\documentclass{aastex62}

\usepackage{amsmath}
\usepackage{stackengine}

\newcommand{\be}{\begin{eqnarray}}
\newcommand{\ee}{\end{eqnarray}}
\newcommand{\Msun}{\mbox{M$_{\odot}$}}

\begin{document}





\title{Outer solar system possibly shaped by a stellar fly-by}

\correspondingauthor{Susanne Pfalzner}
\email{spfalzner@mpifr.de}

  \author{Susanne Pfalzner}
  \affiliation{Max-Planck-Institut f\"ur Radioastronomie, Auf dem H\"ugel 69, 53121 Bonn, Germany}

  \author{Asmita Bhandare}
  \affiliation{Max-Planck-Institut f\"ur Radioastronomie, Auf dem H\"ugel 69, 53121 Bonn, Germany}
  \affiliation{Max-Planck-Institut f\"ur Astronomy, Heidelberg, Germany}

 \author{Kirsten Vincke}
  \affiliation{Max-Planck-Institut f\"ur Radioastronomie, Auf dem H\"ugel 69, 53121 Bonn, Germany}

 \author{Pedro Lacerda}
  \affiliation{Queen's University, Belfast, UK}

\begin{abstract}
The planets of our solar system formed from a gas-dust disk. However, there are some properties of the solar system that are peculiar in this context. First, the cumulative mass of all objects beyond Neptune (TNOs) is only a fraction of what one would expect. Second, unlike the planets themselves, the TNOs do not orbit on coplanar, circular orbits around the Sun, but move mostly on inclined, eccentric orbits and are distributed in a complex way. This implies that some process restructured the outer solar system after its formation. However, some of TNOs, referred to as Sednoids, move outside the zone of influence of the planets. Thus external forces must have played an important part in the restructuring of the outer solar system.   The study presented here shows that a close fly-by of a neighbouring star can simultaneously  lead to the observed lower mass density outside 30 AU and excite the TNOs onto eccentric, inclined orbits, including the family of Sednoids. In the past it was estimated that such close fly-bys are rare during the relevant development stage. However, our numerical simulations show that such  a scenario is much more likely than previously anticipated. A fly-by  also naturally explains the puzzling fact that Neptune has a higher mass than Uranus. Our simulations suggest that many additional Sednoids at high inclinations still await discovery, perhaps including bodies like the postulated planet X.\\
\end{abstract}

\section{Introduction}

Explaining the shape of the solar system once seemed simple -  the planets formed from a smooth disc surrounding the young Sun, orbiting in a plane on almost circular orbits.  However, there are some features of the solar system that seem at odds with this picture. First, the surface density of the solar system drops by a factor of more than \mbox {1 000} outside Neptune's orbit at 30 AU \citep{morbi:03}. Second, most objects outside Neptune (transneptunian objects - TNOs) move on eccentric, inclined orbits (i $>$ 4$^\circ$) relative to the planetary plane \citep{Gladman:2008}. Third, such objects exist even outside the range of influence of the planets. All three points strongly indicate that the outer reaches of the solar system must have been considerably modified by some process(es) that took place after its formation. 

First, we want to have a closer look at the objects that move on peculiar orbits despite being outside the range of influence of Neptune. They are referred to as Sedna-like objects or Sednoids named after Sedna, which was the first observed object of this kind \citep{brown:04}. When another objects of this kind, namely  2012VP$_{113}$, was discovered  \citep{trujillo:14}, it became clear that a population of planetesimals exits that share similar orbital elements  \citep[see also][]{fuente:14}. They are part of the outer regions of the Solar system located between the Kuiper belt and the Oort cloud and orbit on extraordinary wide ($a >$ 150 AU), eccentric orbits with large perihelion distances, $ q >$ 30 AU. All these objects have inclinations with respect to the ecliptic  \mbox{$i$ = 10$^\circ$ –- 30$^\circ$} and arguments of perihelion $\omega$ = 340$^\circ \pm$ 55$^\circ$, so that a common origin has immediately been suggested  \citep{trujillo:14,fuente:14}. By now it is believed that about 20 Sednoids are known, they consist of those originally listed in  \citet{trujillo:14} and  \citet{fuente:14},  additional objects suggested in \citet{sheppard:16} and the recently discovered 2014 UZ224 \citep{gerdes:17}. 

At such large distances from the Sun the Sednoids are outside the area of influence of even Neptune. However, it seems also that their aphelion distances are too short for them to be Oort cloud objects \citep{brasser:15}. In principle,  it would be possible that they were excited to these orbits due to  chaotic diffusion. However, this is also unlikely, because it would require a time longer than the age of the solar system to obtain the distance of Sedna   \citep{sussman:88}. This means that an outside influence is likely responsible for their orbits. There have been several suggestions regarding what such an outside influence could be. One suggestion is that the Sednoids were capture from the outer disc of another star during a close fly-by  \citep{morbi:04, kenyon:04,jilkova:15}. Alternatively, the postulation of a possible wide-orbit ninth planet  or planet X \citep{brown:04,gomes:06,soares:13,batygin:16} has gained renewed popularity recently.

Here we suggest a somewhat different process whereby the Sednoids would have been excited to their current orbits also due to the close fly-by of a star. However, in contrast to the capture scenario, here the Sednoids would originate from the Sun's own, originally more extended disc. In section 2 we test in how far one can obtain the different properties of the TNOs, namely, the Senoids, the cut-off at 30-35 AU, the Kuiper belt population and some additional outer solar system features. Afterwards we restrict the parameter space to fly-bys that match best all the outer solar system properties. This is done by performing an extensive parameter study of the effect of fly-bys on discs and applying the criteria that correspond to the observed properties to the obtained simulation results. We will see that the beauty of this scenario lies in the fact that one can obtain all these features, and more, in one single event. 

A new theory has not only to pass the test of reproducing the data and being fairly simply, which this model fulfils, but also that such a process is likely to happen.  Therefore we perform simulations to determine the frequency of such close fly-bys in star clusters as the birth environmemt of the solar system. Here we model the different phases of star cluster development and make predictions for each of these periods in section 3. We discuss the various options when such a fly-by could have happened during the past 4.65 Gyr since the formation of the Sun.  In section 4 the results are detailed and compared to other models for shaping the outer solar system.

\section{Relevant fly-by parameter space}

\subsection{Method}
 In a first step we determine the parameter space of fly-bys that leads to a cut-off at 30-35 AU. We simulate the Sun as being surrounded by a disc of test particles. This disc could represent either a protoplanetary or a debris disc, the inner part may even contain the already formed planets. Accordingly, we assume that the disc mass $m_\mathrm{disc}$ is much smaller than the mass of the Sun, which is the case for most protoplanetary discs and all debris discs. Under these conditions, one can neglect viscosity and self-gravity between the disc particles, unless one is interested in the long-term behaviour after the fly-by. Although the giants planets, in particular Neptune, influence also the orbital parameters of the TNOs over the Gyr, we do not include these forces explicitely in our simulations, because in this first proof-of-principle study we are mainly interested in the properties of distant TNOs, like Sedna, just after the fly-by. On these distant TNOs the influence of the planets is negligible (for more details see section 2.2 and Eq. 3 therein). In this case the problem reduces to a gravitational three-body interactions between the Sun, the perturber star and each disc particle \citep{hall:96,kobayashi:01,musielak:14}. 

\begin{table}[b]
  \begin{center}
    \caption{Parameter space of the modelled fly-bys. The pertruber mass is given in solar masses, the periastron distances in AU and the inclination and orientation in degrees.}
    \begin{tabular}[t]{ll} \tableline \tableline
Parameter & Simulated values\\[0.5ex]
\hline
perturber mass        &0.1, 0.3, 0.5, 1.0, 2.0 5.0, 10. 20, 50\\  
periastron distance & 30, 50, 100, 120, 150, 200, 250, 300, 500, 700, 1000\\
inclination               & 0,10,20,30,40,50,60,70,80,90,100,110,120,130,140,150,160,170,180\\
periastron angle & 0, 45, 90\\
\tableline
    \end{tabular}
     \label{tab:set-up_params}
  \end{center}
\end{table}

Discs with an outer radius of 100 AU, 150 AU or 200 AU were modelled with the tracer particles initially orbiting the Sun on circular Keplerian orbits. An idealized thin disc \citep{pringle:81} was assumed with all particles initially located on the same plane (but see exceptions in section 2.2). In our parameter study we used 10 000 mass-less tracer particles to model this disc. This particle number is a compromise made to be able to scan the extensive parameter space while still resolving the relevant populations. For the here considered problem it is relevant to have a sufficiently high resolution in the outer part of the disc. This was achieved here by using an initially constant particle surface density and post-processing the data by assigning different masses to the particles to model the actual mass density distribution \citep{hall:96,steinhausen:12,ovelar:12}. This is computationally much more efficient than increasing the particle number. 

 We modelled fly-bys with perturber masses of 0.3, 0.5, 1.0, 2.0 5.0, 10. 20, 50 \Msun\ at  periastron distances of $r_\mathrm{peri}$= 30, 50, 100, 120, 150, 200, 250, 300, 500, 700, and  1000 AU.  The parameter space in orbital inclination was covered in steps of 10$^\circ$ and the angle of periastron in steps of 45$^\circ$. The extent of our parameter study is summarized in \mbox{Table 1,} amounting in total to 5643 parameter combinations. It is assumed that the perturber moved on a parabolic orbit, which is a reasonable assumption as long as the star cluster is not extremely dense. As we consider here a cluster environment similar to Orion nebula cluster (ONC) this assumption is justified  \citep{olczak:10,olczak:12,vincke:16}. 

A Runge-Kutta Cash-Karp scheme was used to determine the particle trajectories. The simulation starts and ends when it holds for all particles bound to the Sun that the force of the star flying by is less than  0.1\% of that of the Sun. 
 To obtain a statistically sound sample we performed 20 simulations for each fly-by scenario with different random seeds for the initial particle distribution.  At the end of the simulation the properties like eccentricity, inclination, periastron distance, semi-major axis, orbit averaged position etc. are stored for each test particle. For a more detailed description of the numerical method see \citet{bhandare:16}.  The resulting particle distributions are then compared to the observational data.

 \subsection{Fly-by processes leading to solar system-like features}
First we select a subset of simulations that yield  a disc size in the range of 30-35 AU. In the literature one often finds that the disc size is reduced to 0.3-0.5 times the periastron distance. However, this approximations holds only for the fly-by of a solar mass stars \citep{kobayashi:01}. In a cluster a wide variety of stars with masses ranging from 0.08 \Msun\ up to $\approx$  100 \Msun\ might in principle act as  a perturber. However, not each perturber mass is equally likely. Actually fly-bys of stars with mass $\approx$  0.5 \Msun\ or stars with masses exceeding 5 \Msun\ are the most common ones to lead to disc truncation \citep{olczak:10, vincke:15}. This is at least so for clusters like the Orion nebula cluster (ONC), that we will discuss in more detail in section 3.2. The reason is that  0.5 \Msun\ stars are the most common type of star according to the initial mass function, whereas massive star function as gravitational foci in the clusters.  
However, to cover the entire parameter space of possible fly-by masses the relation
\begin{figure}[t]
\includegraphics [scale=0.15] {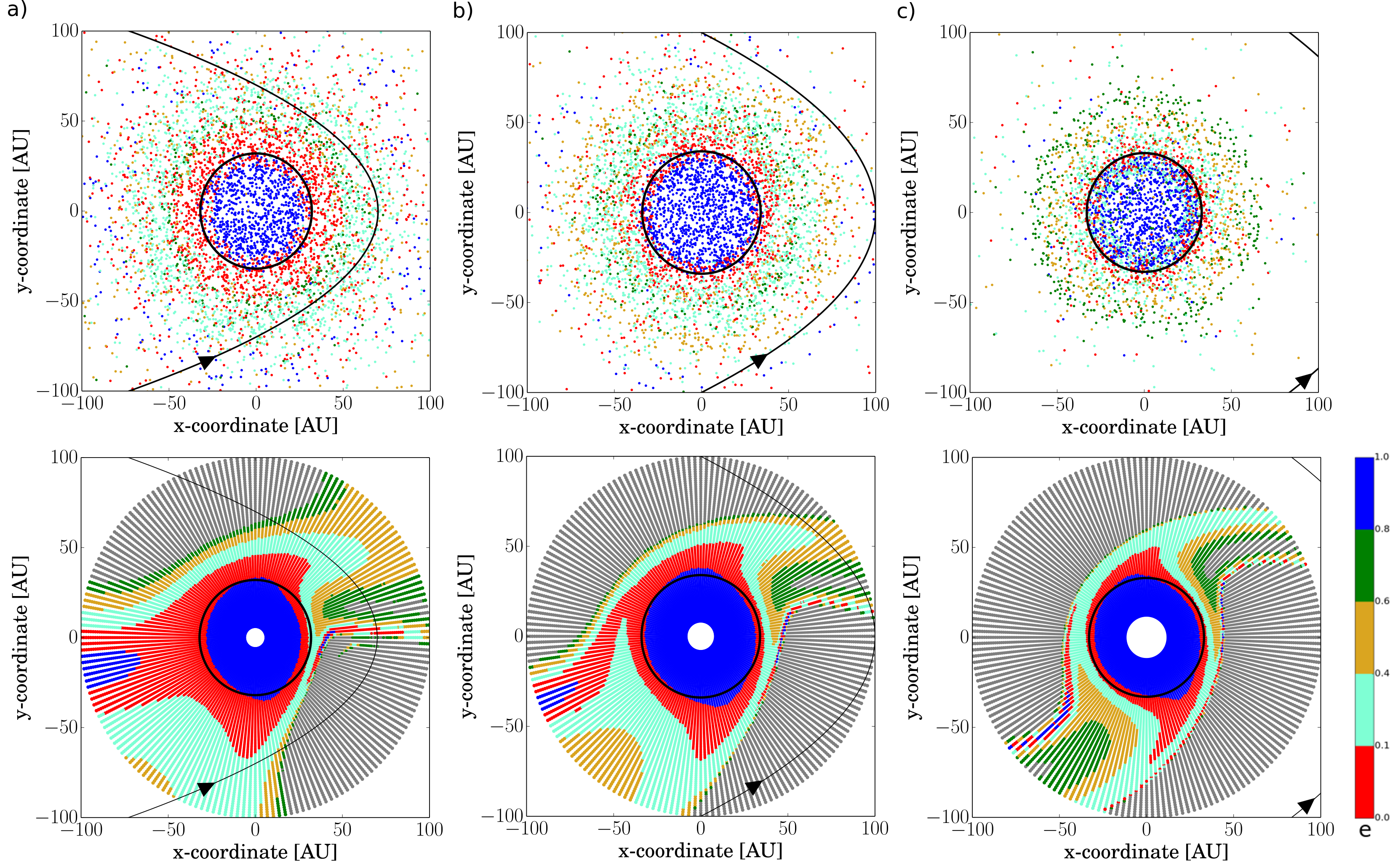}
\caption{Effect of a prograde, parabolic fly-by of a star with a) $M$=0.5 \Msun, b) $M_2$= 1, \Msun  and c) $M_2$= 5 \Msun that is inclined by 60 degree and has a angle of periastron equal zero. The perihelion distance is always chosen in such a way as to lead to a 30-35 AU disc. The top row indicates the eccentricity distribution of the matter with a central area of most particles on circular orbits and more eccentric orbits at larger distances form the Sun. The eccentricities are indicated by the different colours given in the bar. The origin of the different eccentricity populations in the original disc can be seen in bottom row, where matter indicated in grey becomes unbound from the Sun. Note that in c) the path of the perturber is not visible because it is outside the shown frame.}
\label{fig:size:regime}
\end{figure}
\be r_d = 0.28 \times M_p^{-0.32} r_\mathrm{peri}\ee
given by \cite{breslau:14} points at the relevant parameter space of fly-bys that that lead to a 30-35 AU drop-off. 
With $r_d$ = 30-35 AU this leads to
\be   r_\mathrm{peri} = (107 - 125) \cdot M_p^{0.32} \ee
for the periastron distances $r_\mathrm{peri}$ that leads to a solar system sized disc when a star of mass $M_p$  passes.
Strictly this relation is only valid for the DPS size after coplanar fly-bys, but as we model anyway the entire parameter space given in Table 1 it functions as a first indicator to determine the relevant subset of fly-bys. 

In a next step we identify in this subset the fly-bys that lead simultaneously to a population that represents the Kuiper belt reasonably well and produces Sednoids. Here one has to keep in mind that after the fly-by the objects in the inner Kuiper belt region will be affected by interactions with Neptune for a considerable time afterwards. We do not model this long-term behaviour here. Therefore it is not our aim to reproduce the inner Kuiper belt structure as closely as possible. Instead our prime goal is to obtain a good match for the Sedna-like objects, while having a reasonable fit for the Kuiper belt. 

Fig. 1 illustrates fly-bys of perturbers of different mass  \mbox{($M_2$=0.5, 1, 5 \Msun )} that lead to a mass density drop at 30-35 AU distance from the Sun.
The top row shows the path of the perturber star in the rest frame and the time averaged positions of the test particles of the disc. The colour scheme represents the eccentricities of the particles, with grey representing the particles that become unbound from the Sun due to the fly-by. In all three cases it is apparent that a fly-by creates in principle a similar pattern with an inner unperturbed region and outside this area the eccentricity of the particles increase on average with distance to the Sun. The bottom row shows where the particles with the different eccentricities where originally located in the disc (for details of the method see  \citet{breslau:17}.  It can be seen that the fate of the particles follows a complex pattern. 

Despite the general similarity of the three sets of fly-by parameters there are considerable differences in the location of the different regions and the relative portions of particles becoming or having similar properties. In short,  higher mass stars lead to more compact configurations with very few objects beyond 60 AU - considerably less than one finds in the solar system. We find that masses in the range $M_2$=0.5 - 1 \Msun\ usually correspond not only better to the Kuiper belt population than fly-bys of higher mass stars, but are the only ones able to produced objects like Sedna.

However, it is not only the periastron distance and the perturber mass, that influence the resulting population, but also the relative orientation between the plane of the disc and that of the path of the perturber, which is extremely important. In order to obtain a good fit with the inclinations of the TNOs the fly-by has to be inclined and at a certain orientation. Going through the so-obtained subset, we look for those that give the best fit to all three properties - 30 AU drop-off, Kuiper belt and Sednoids. Of the
parameter combinations listed in table 1, a very good match to the observed properties is found for the fly-by of a star of mass $m_p$=0.5 \Msun\ on an inclined orbit (60$^\circ$) with an angle of periastron of 90$^\circ$ passing the Sun at 100 AU, that was initially surrounded by a 150 AU-sized disc. Figure 2a shows the distribution of particles after the fly-by similar to Fig. 1a, but this time on an inclined orbit as described above.   
This fly-by effectively truncates the disc, leaving the system inside 30 AU basically undisturbed (Fig.~2a) while depleting the total mass in the Kuiper belt region to $<$ 1\% of its initial mass. At the same time it excites the trans-Neptunian belt and forms a family of Sednoids.  Fig. 2b illustrates the original positions of the particles color coded with their final eccentricity after the fly-by. Here dark green corresponds to the possible sites of the origin of Sedna. 
During such a fly-by matter is not exclusively transported outwards, but a considerable fraction also moves inwards. The reason for this behaviour is that not only super-Keplerian velocities are induced by fly-bys, but also sub-Keplerian velocities \citep{pfalzner:03}, see Fig. 2c.

\begin{figure}
\includegraphics[scale=0.31]{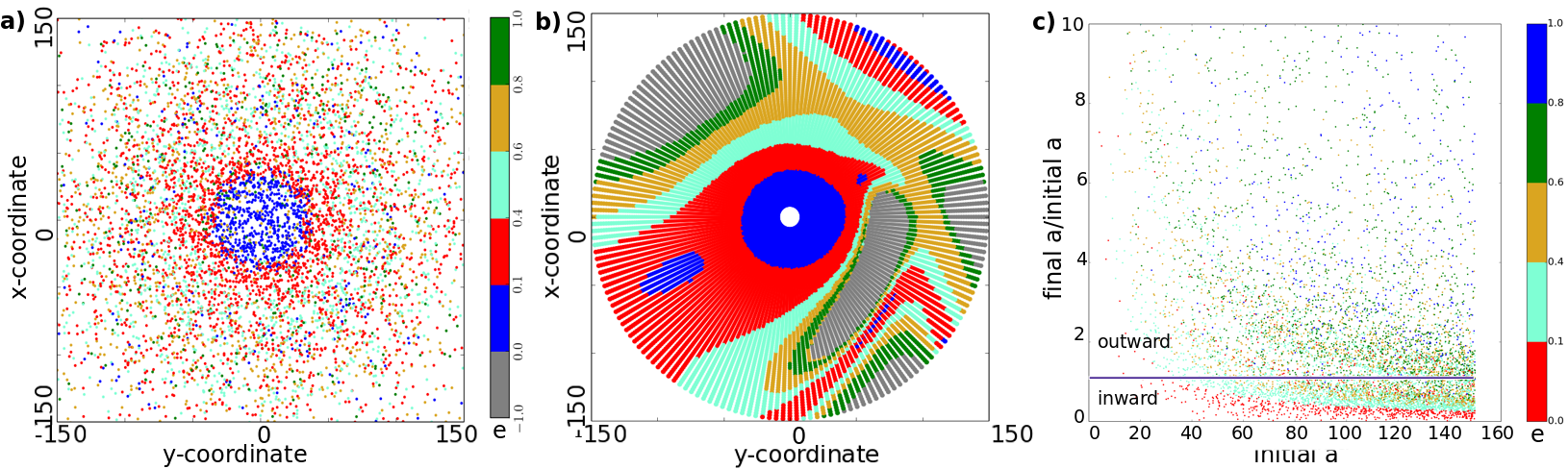}
\caption{Effect of a prograde, parabolic fly-by of a star with $M$=0.5 \Msun, a  perihelion distance of 100 AU at an inclination of 60 degree tilted by  90 degree on a disc or planetary system. a) indicates the eccentricity distribution of the matter with a central area of most particles on circular orbits and more eccentric orbits at larger distances form the Sun. The origin of the different eccentricity populations can be seen in b) and c) shows the relative final distance to the Sun vs. the initial distance. }
\label{fig:example}
\end{figure}

The fact that the fly-by of  a star with mass  $\approx$  0.5 \Msun\ gives a relatively good fit,  is in accordance with above mentioned result that the most common disc truncating fly-bys  in a cluster are those with  $\approx$  0.5 \Msun .  During this fly-by for all planets the induced eccentricity are even in extreme cases always $<$0.05, but can be considerably lower, depending on the relative position of the planets at the moment of periastron passage of the perturber.  Thus the influence on the planetary orbits is negligible. 

The fly-by parameters we obtain are very similar to the case  \cite{kobayashi:05} investigated for the formation of the Kuiper belt population, where they modelled also the effect of 0.5 \Msun\ perturber. However, they did not obtain the cold Kuiper belt population nor a Sedna-like object.  \cite{punzo:14} had a similar problem for different fly-by parameters. In both cases the reason is two prime differences in the set-up, namely, we have a considerably higher resolution in the outer regions of the disc and our initial disc is larger. Little as these differences seem they have severe consequences. In smaller discs one obtains many of the other Senoids,  but not Sedna itself because its point of origin is most likely  between 100 and 120 AU. The absence of a population with low eccentrcities (cold Kuiper belt) in these simulations is because this population is considerably smaller than that of the hot Kuiper belt. Thus one requires a relatively high number of test particles in the disc outskirts, which is more difficult to achieve with a 1/r resolution in the test particle distribution.

Before discussing the different populations that can be seen in our simulations, we want to point out, that although above given parameters lead to a relatively good fit, it is far from clear whether they are the optimal choice.  Due to the extensive parameter space, our grid of simulations is  necessarily still quite coarse. It is better to say that if  a fly-by really shaped the outer solar system, then it would have been by a star in the mass range  $M$=0.3 - 1.0 \Msun\ passing at a perihelion distance of between 50 and 150 AU, with an inclination in the range 50 to 70 degree and an angle of periastron between 60 to 120 degree. The next steps would be to test with a finer grid in the relevant parameter space and calculate how the Kuiper belt population would develop after a fly-by due to the interactions with Neptune. Both tasks are beyond the scope of this work because (i) the latter would require a different type of code again and (ii) one would need still better observational constraints on the Kuiper belt and Sednoid population to have a solid case to test for in an automated way.

\begin{figure}
\includegraphics[scale=0.46]{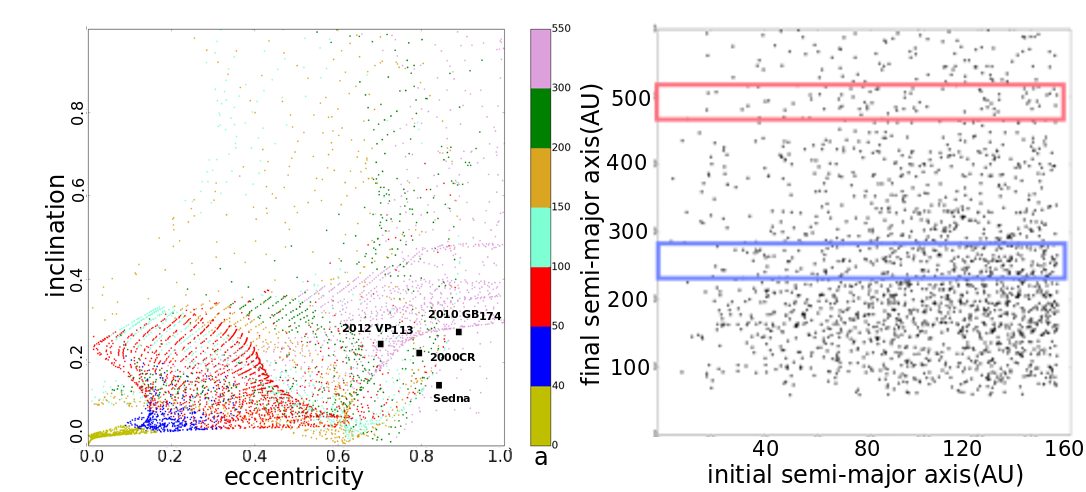}
\caption{a) shows the eccentricity vs. inclination for all test particles for the fly-by illustrated in Fig. 2, with the properties of Sedna, 2012 VP$_{113}$, 2000 CR and 2010 GB$_{117}$ indicated. b) shows the  semi-major axis after the fly-by (final) of the test particles as a function of the pre-fly-by semi-major axis (initial). The areas relevant for Sedna (red) and  2012 VP$_{113}$ (blue) are high-lighted. }
\label{fig:example}
\end{figure}

Fig. 3 shows the eccentricity vs. inclination distribution after the fly-by with above selected properties. For comparison we included the four most extreme Sednoids in the plot. It can be seen that Sednoids would be a natural outcome of such a fly-by.   To obtain a closer grip onto where the Sednoids would have most likely resided before the fly-by, we look in \mbox{Fig. 3 b} where particles with final semi-major axis equivalent to that of Sedna (red) and 2012 VP$_{113}$ originate from.
It can be seen that in principle any location between 30 and 150 AU would be possible, but that the likelihood Sedna and 2012 VP$_{113}$ originating from the area between 100  tp 150 AU is considerably higher.  This explains also why neither  \cite{kobayashi:05} nor \cite{punzo:14}  found Sedna-like objects, as they modelled only populations reaching out to 100 AU and 90 AU, respectively.  We see in Fig. 4c that in the  eccentricity vs. periastron plane also Sedna-like objects are clearly expected after such a fly-by. In summary, Sedna-like objects are a common outcome of fly-bys under the condition that the initial disc extended at least to 100-120 AU.

Next, we present a detailed comparison of the TNO population expected from our simulations with the observed one. Figure 4 compares known TNOs to simulation particles. To ensure a fair and relevant comparison, we apply two criteria: a) we plot only TNOs and simulation particles that are not strongly coupled to Neptune, and b) we plot only simulation particles that would likely be detected from Earth. 

Firstly, we apply criterion a) by selecting only TNOs with $a>30$ AU and $T_N>3.05$. $T_N$ is the Tisserand parameter with respect to Neptune, given as

\begin{equation}
T_N = \frac{a_N}{a} + 2 \left[(1 - e^2)\frac{a}{a_N}\right]^{1/2} \cos(i)
\end{equation}

\noindent where $a_N$ is Neptune's semimajor axis and $a,e,i$ are the semimajor axis, eccentricity and inclination of the TNO. The Tisserand parameter is commonly used to distinguish asteroids from Jupiter-family comets in terms of their dynamics with respect to Jupiter: asteroids, with $T_J>3.05$, are dynamically decoupled from Jupiter, whereas comets, with $T_J<3$, have orbits that are strongly coupled to the gas giant \citep{jewitt:09}. Analogously, we use $T_N>3.05$ to select only those TNOs that are dynamically decoupled from Neptune. We apply similar dynamical criteria to the model particles.

Secondly, we impose the detectability threshold b) to the model particles. TNO surveys have complicated detection biases, but the main and most general effect is that TNOs tend to be discovered when at perihelion \citep{shankman:17}. To account for this, we select only simulation particles with perihelia less than $q=80.4$ AU, which is the value for 2012 VP$_{113}$, the TNO with the largest known perihelion \citep{trujillo:14}.

 \begin{figure}
\includegraphics [angle=0,scale=1.2] {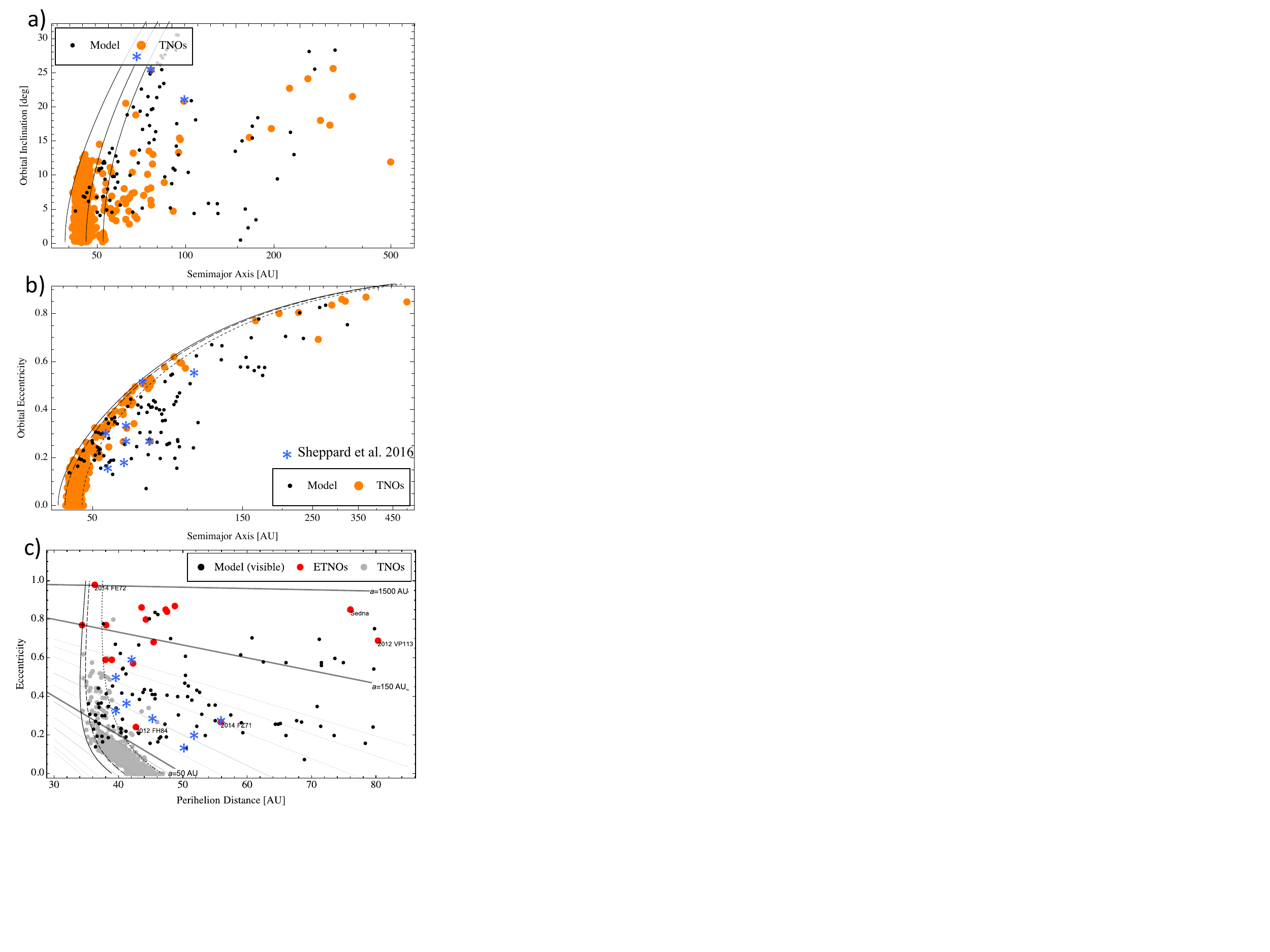}
\caption{Resulting distribution of objects after the fly-by, where a) shows inclination vs. semi-major axis, b) eccentricity vs. semi-major axis and c) eccentricity vs. perihelion distance. The red symbols indicate the observed objects and the black symbols the predictions from the model.}
\label{fig:size:regime}
\end{figure}

 As already pointed out by  \cite{kobayashi:05} and \cite{punzo:14} fly-bys can produce the excited orbits dominant in this region very well.  This can be seen in Fig. 4 were the simulation results are compared with the observed properties of the TNOs.  Here the representation of the simulation results is different to those in the previous plots as we assumed a $1/r$-dependence of the initial disc and weight accordingly. The particles in the plot are only those with a high mass density in this area. We chose this approach because this probably resembles the observational selection effects. However, our results indicate that this may be problematic for some features of the TNO population. We see in accordance with  \cite{kobayashi:05} and \cite{punzo:14} that the overall properties of the hot Kuiper belt are well reproduced. 

However, their problem was that the cold Kuiper belt population was missing, they had no particles with $e <$ 0.1. We pointed out earlier that for the work by \cite{kobayashi:05} this was most likely a resolution problem. If we look at our high resolution plot (see Fig. 3), we clearly see a population with the eccentricity all the way down to $e$=0. This can be seen better in Fig.5, where we explicitly show only this population. (Note, this area has a relative low mass density, which is why it is missing in Fig. 4a, where we show only areas of high mass density.) This confirms that one needs a sufficiently high resolution to obtain the cold Kuiper belt population.

Going back to Fig. 3b we can see where this low eccentricity population originates from. All matter on fairly circular orbits are indicated in blue in this plot. Interestingly we find several regions outside the central planet area where the particles still move on orbits with $\epsilon <$ 0.1 after the fly-by, some of them at very large radial distances ($>$ 100 AU).  In addition, there is a small population scattered within the red area (0.1 $< \epsilon <$ 0.4). In our scenario the cold Kuiper belt population would originate from this population plus the small blue area at 45-50 AU. Basically, it is the population more or less, at the opposite side of the periastron passage that remains fairly undisturbed. The relative size and location of this area depends strongly on the actual fly-by parameters. This difference is highlighted when compared to the equivalent coplanar fly-by (Fig. 1a, bottom), where these areas (here in red) are at very different locations and much smaller. Therefore it is no longer the question whether a fly-by gives also a cold Kuiper belt population, but which fly-by parameters give the best match to the observed population.

\begin{figure}
\centering
\includegraphics[scale=0.3]{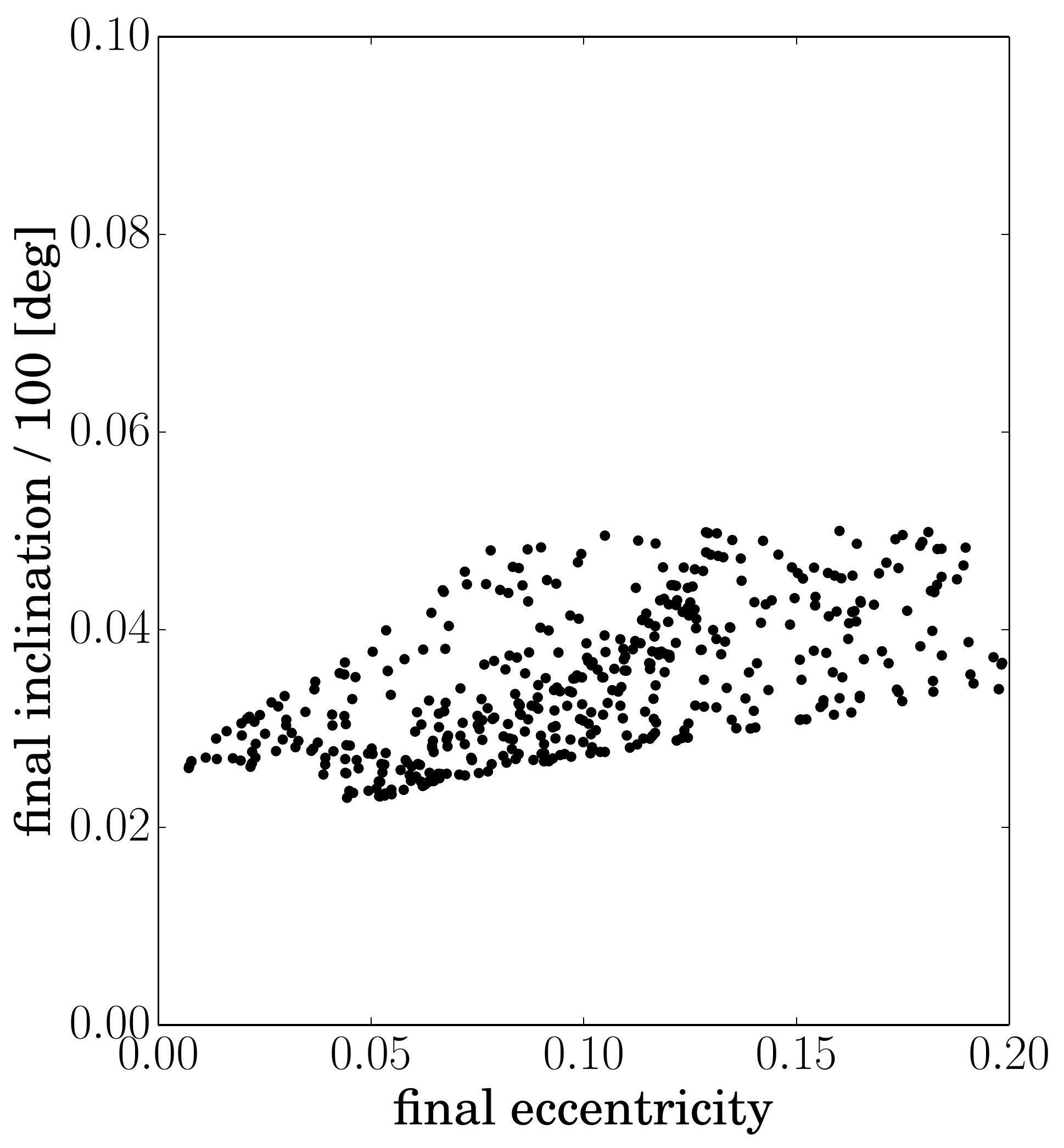}
\caption{Close-up of the cold Kuiper belt population from Fig. 3a.}
\label{fig:example}
\end{figure}

Fig. 3a also shows that there is an actual gap between the cold and hot population in inclinations similar to the observed one. 
However, for our scenario we find a slightly lower number of cold Kuiper belt objects (5\%-8\% of the total Kuiper belt population) compared to the observed $\sim$ 10\%.
In addition, most objects are more inclined than in the observed population. 
The low number of cold Kuiper belt objects and the higher inclination may either be due to the here missing long-term evolution or the model assumptions. 
Here we assumed an idealized thin disc which might not be realistic. If the disc had a non-zero scale height, the number of low inclination objects among the cold Kuiper belt population would probably be higher. Second, \citet{punzo:14}  showed that secular evolution quickly leads to `cooling' of the eccentricities and inclinations making the  population colder. The importance of both these effects will be investigated in follow-up studies. 

Closer inspection reveals some additional features in Fig. 4. \cite{sheppard:16} found a new family TNOs which have relatively large periastra but low eccentricities. These are also present in our, so far, best fit solution. We have indicated these objects in Fig. 4 as blue asterisks. Actually we would expect that similar objects but with even larger periasta (100-200 AU) would await detection.

We must admit, that there is also one feature which we would expect from our simulations, but has so far no equivalent in the observations. That is a population with large periastra (100-200 AU), large eccentricities (0.6-0-8), but low inclinations (5-10 degree). We will have to check in the follow up study whether this population always occurs in such fly-bys or is just a feature of the particular parameters we chose here. 

\begin{figure}
\centering
\includegraphics[scale=1.0]{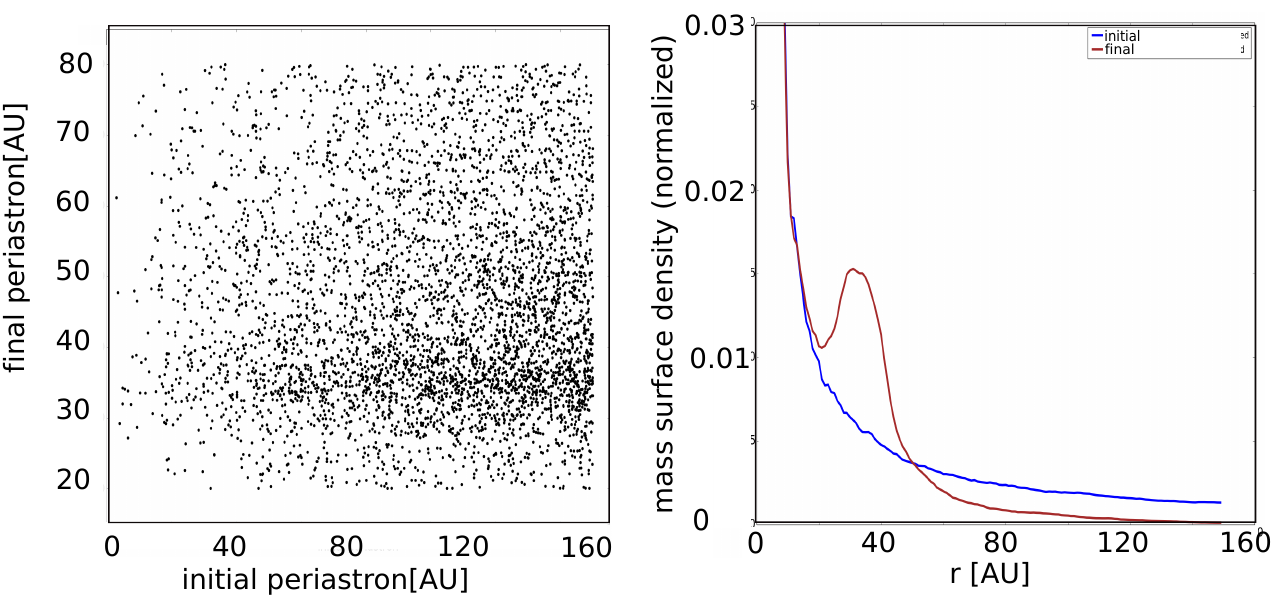}
\caption{fFnal vs. initial distance to the Sun of all test particles.}
\label{fig:example}
\end{figure}

Unlike the so far detected exoplanetary systems, where all planets of a system seem to be of fairly equal mass and seperation, the seperations of the giant planets in the solar system increases while their masses decreases steadily with distance to the Sun\citep{weiss:18}. The mass distribtion of the solar system planets is usually attributed as being a direct consequence of that in the disc they formed from.
There is only one exception,  Neptune's mass (17.2 $M_\mathrm{Earth}$) is larger that that of Uranus(14.6 $M_\mathrm{Earth}$). There have been several theories trying to explain this situation \citep{levison:08}, however, a fly-by would also naturally explain this fact. We saw earlier that as a consequence of the fly-by most matter beyond Neptune's orbit moves outwards, but some portion also moves inwards. 
This matter comes partly from very distant locations (50 - 150 AU) and concentrates somewhere between 30 and 35 AU (see Fig.6 a). This matter transport creates a bump which is located at 35-60AU right after the fly-by. Afterwards, it moves inwards (Fig. 2c) and the particles have final eccentricities of 0.0-0.4 (see Fig.~2c, bottom). The matter gets accreted every time the particles are at their perihelion and approach closely Neptune's orbit. With the entire matter contained in this "bump" corresponding to approximately 1-2 Earth masses, this is somewhat less than the 2.6 $M_{Earth}$ mass difference between Uranus and Neptune, but self-gravity effects or slightly different fly-by parameters could account for this difference. This will be tested in follow-up studies.


In summary, a fly-by with the parameters in the above described range reproduces many of the features of the outer solar system all in one go.  Thus this model fulfils two 
demands on a new theory - it agrees largely with the available data and it is simpler than existing models. The question is now: Is it also likely that such an event happened?
In the following section we want to address the question of how likely such close fly-by event would have been during the last 4.56 Gyr.

\section{Frequency of solar systems-forming fly-bys}

Fy-bys are often regarded as quite rare events, but they happen even today, however, most of them are distant and mainly influence the outer Oort cloud \citep{bailer:18}. Nevertheless, they provide a considerable amount of long-period comets \citep{mamajek:15,feng:15,dones:15,feng:17}.  The closest recorded fly-by in recent times, was that of  WISE J072003.20-084651.2 (“Scholz star”) which passed approximately 70 000 years ago at a distance of $\approx$ 14 000 AU \citep{mamajek:15}. Thus it was 140 times more distant than the type of fly-by we are considering here.
However, as the Sun formed like most sta´rs in a cluster environment the close fly-by frequency is generally assumed to have been considerably higher in its initial stages.
Existing estimates of close fly-by frequencies are mostly based on the simulations of the fly-by rate in a cluster with a density distribution similar to the Orion Nebular cluster\citep{pros:09,adams:10}.  In  contrast to the here modelled temporal evolution of the cluster, they investigate the case of a relatively constant average stellar density of 100 stars/pc$^3$. They find that the frequency of fly-bys at 90 AU should occur with a rate of 10$^{-4}$ -10$^{-2}$ Myr$^{-1}$ during the first 10 Myr. 
Their argument continues with the frequency value 10$^{-3}$  Myr$^{-1}$ and they conclude that this is a low probability.

Given that 90\% of the Milky way clusters largely dissolve within 10 Myr \citep{lada:03}, it is often assumed that basically no close fly-bys happen afterwards.
This is the main reason why previous suggestions of fly-bys possibly being responsible for the properties of the outer solar system have received not much attention.
It is often argued that in these outer areas objects would require well over 10 Myr to grow to their current size and that at that time fly-bys would be extremely rare. 
However, as already suggested by \citet{davies:14}, despite the fly-by frequency at later times being relatively low, even at such a low occurence rate there was a realistic change for close fly-by given the long timespan of 4.5 Gyr. They estimated that a close fly-by is approximate as likely to have happened during the first 10 Myr as during the consecutive 4.5 Gyr.  

Here we want to have a renewed look at the frequency of close fly-bys for the following reasons:
\begin{itemize}
\item Newly available high-resolution images of discs around stars younger than 10 Myr show prominent ring structures at several tens to hundreds of AU. Many authors interpret these as signatures of already formed or currently forming planets \citep{testi:15,andrews:16,fedele:18}. If gas giants can form at such distances from their host star in such a short time span, it can no longer be excluded that TNO-sized objects could form on time scales $<$ 10 Myr. This means the argument against an early fly-by  ($<$ 10 Myr) has become weaker. 
\item The estimate of the fly-by frequency made several simplifying assumptions. First, a cluster consists of a broad spectrum of stellar masses, taking an equal-mass fly-by as example might not be representative. Second, the cluster was assumed to be in a quasi-steady state, meaning the stellar density remained approximately constant over the simulation time. However, clusters are highly dynamical systems, so it might be essential to determine the fly-by frequency during all phases of the cluster development, from the embedded phase, the gas expulsion stage to the expansion phase and beyond. Third, clusters exist with extremely different stellar densities, the ONC might not be representative for the solar birth cluster.
\end{itemize}

The question is how often do close fly-bys that lead to a significant drop in mass density at 30 AU actually happen. This obviously depends on the stellar density in the cluster.  Although it seems fairly certain that the Sun formed in a cluster environment, it is still unclear what type of cluster it was\citep{adams:14,pfalzner:15}. Star clusters vary widely in masses and sizes.  Here we present our results for a cluster similar to the Orion nebula cluster (ONC), because it has been used as example in previous studies. However, for more compact clusters, like NGC 3603, the fly-by frequency could be considerably higher due to their higher stellar density. Thus the here presented values can most likely regarded as lower limits for the close fly-by frequency.  However, the density development in such  massive clusters are well constrained by observations \citep{pfalzner:09,pfalzner:13}. In Vincke \& Pfalzner (in prep.) we will give the close fly-by frequency for a variety of clusters, which will put the here presented values in perspective. 

\subsection{Method}

Here only a short summary of the method is given, details including a discussion of the assumptions are given in \citet{vincke:16}. The simulations presented here are equivalent to those of model E2 in \citep{vincke:16}. The clusters was modelled using 4000 simulation particles, corresponding to the estimated number of stars in the ONC.  We assumed an initial cluster half-mass radius of \mbox{$r_{hm} = 1.3$pc} with the initial stellar number density distribution being a relaxed King distribution (W9). For simplicity  we neglect potentially existing initial sub-structuring of the cluster \citep{parker:12,craig:13} in this study. Such sub-structuring could potentially increase the initial rate of close fly-bys \citep{winter:18}, so that again our estimate will be quite conservative. However, after the first 2-3 Myr substructuring should be largely erased. 

One of the main differences in comparison to other simulations is that 
we follow the dynamical evolution of such a clusters from the point where star formation has finished, through the gas expulsion phase resulting in cluster expansion until the cluster reaches an age of 10 Myr.  The simulations were performed using the code Nbody6GPU \citep{aarseth:03}. The cluster was assumed to be in virial equilibrium initially. A star formation efficiency of  30\% was modelled with the gas expulsion taking place instantaneously after 2 Myr.   The  fly-by history of the first 10 Myr is recorded, at which stage the clusters have found a new quasi-equilibrium state after gas expulsion.
Afterwards there is some cluster mass loss due to ejections and stellar evoltion taking place, but this decreases the cluster density by no more than a factor two. Therefore the fly-by rate at 10 Myr will be fairly characteristic at least during the consecutive 1 Gyr.

Several simulations with different seeds in the set up were performed to improve the statistical significance of the results.  In practice we start with every star being surrounded by a disc with a size of 500 AU, but any other size value would do as long as it is larger than the 30-50 AU that we test for. In each simulation, the fly-by history for every individual star is tracked and the relation given by eq. 2 is then used to determine the system size after coplanar, prograde, parabolic fly-bys. This information is then used to determine the frequency of fly-bys that lead to a disc size of 30-50AU for solar-type stars, which is here defined as stars in the mass range 0.8\Msun\ $\leq M \leq $1.2\Msun . Inclined fly-bys lead to slightly larger disc sizes than coplanar fly-bys \citep[][]{clarke:93, heller:95,bhandare:16}, but for a first estimate of fly-by frequency eq. 1 should suffice.

\begin{figure}
\centering
\includegraphics [angle=0,scale=0.5] {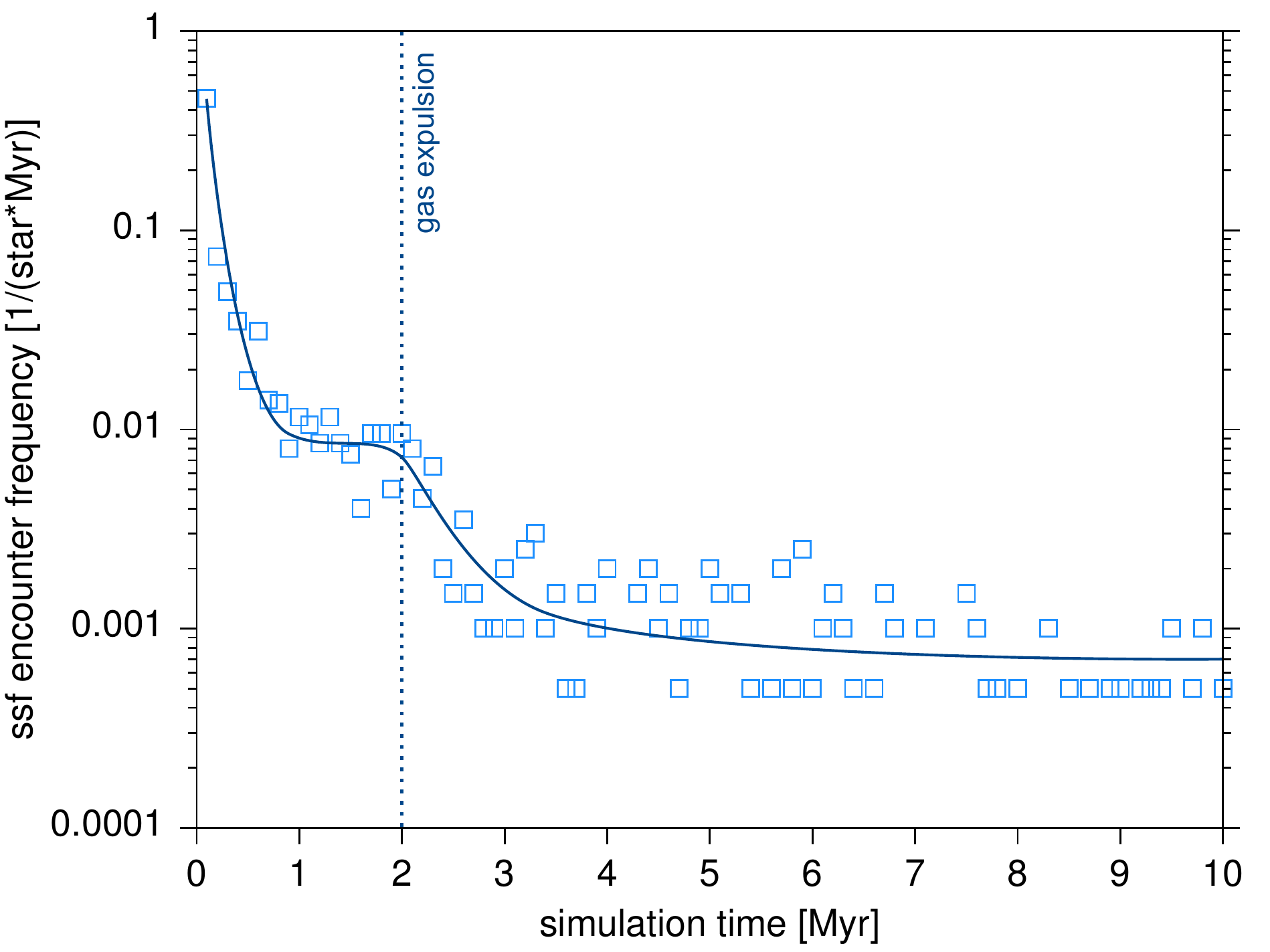}
\caption{Fly-bys frequency as a function of time in an ONC-like cluster. This refers to fly-bys that lead a disc size of 30-50AU for solar-type stars. Gas expulsion is assumed to be instantaneous and take place at $t$=2Myr.}
\label{fig:size:regime}
\end{figure}

\subsection{Results for the fly-by frequency}

During the first 10 Myr  of cluster development the average stellar density decreases about 3 orders of magnitude in the cluster. The decrease in stellar density is most dramatic just after gas expulsion, which was assumed to happen at 2 Myr. The decrease in stellar density translates into a drop in the probability of a fly-by that truncates the disc size to the solar system size. This is illustrated in Fig. 7 which shows the probability of such a fly-by for solar-type stars. The probability for such a fly-by is highest during the first Myr where it is above 10\% in the very first few  \mbox{100 000 yr} but drops to a rate of 1\%, where it remains during the embedded phase.  Note, that the value at 1 Myr corresponds roughly to the high end of the frequency estimated by \cite{adams:10}. During the first Myr 3-5\% of solar type stars would experience such a fly-by in an ONC-like cluster, and the chance for a close fly-by would increase to about 5-7\% chance over the first 10 Myr. As such solar system shaping fly-bys are relatively common during that phase. Thus, if Sedna-like objects can form on such short time scale as some interpretations of the ring structure suggest, a solar system forming fly-by would be a relatively frequent event.

However, if Sedna-sized objects need considerably longer than just \mbox{1-2 Myr} to grow,  say at least 5 Myr, the question arises whether the dramatically decreased stellar density after gas expulsion would still allow for such an event being likely to happen. Fig. 7 shows that after 5 Myr the fly-by frequency remains more or less constant. Here we take the fly-by probability at 10 Myr as a guideline, because at this age the clusters have reached their new semi-equilibrium state and their density declines only very slowly afterwards. 
Fig. 7 shows that the fly-by probability at 10 Myr  lies at approximately 0.7-0.8 \% per Myr. From long-term simulations of clusters it is known that the density in remnant clusters decrease about a factor 2-3 on a Gyr time scale. This means that the fly-by frequency would decrease by the same factor. However, the Sun exists since 4.56 Gyr, so even during the first Gyr every solar-type star would have experienced 0.2-0.3 such events, or in other words there would have been a 20-30\% probability for such a fly-b during the first Gyr alone. Thus it is actually  likely for a solar-type star to be exposed to such a close fly-by in the long time span that passed since its formation. of




One obvious question is: when would this fly-by most likely have happened?  
Our simulations confirm the earlier result by \citet{malmberg:11} that  although the fly-by rate is highest in the initial phases, this can often be outbalanced by the long timespan afterwards giving similar probabilities. If we take the values from above, it means that actually the likelihood that the fly-by happened in the later stages was actually a little bit higher than during the first 10 Myr. However, this depends on what type of cluster the Sun actually formed in.
If the Sun indeed formed in an ONC-like cluster, a prime candidate for the fly-by would be the time of the late heavy bombardment, since a fly-by would be an extremely good explanation for the large number of impacts. As we saw above, a fly-by does not only lead to matter being excited to more distant orbits, but some matter is also transported inwards. A small fraction is transported actually to the inner regions (inside 30 AU) of the solar system. However, in how far the properties of this matter corresponds to what one expects from that responsible for the late heavy bombardment will need further investigation.

\section{Comparison with other models}

Next we want to compare the here proposed model with other models that explain outer solar system features. In this comparison we concentrate on the capture scenario and the Nice model. We want to point out that this is not necessarily a question of either/or, but that the capture model and the Nice model are potentially compatible with the scenario we presented above.

\subsection{Nice model}

We start with presenting a short summary of the model often referred to as Nice model (for a more detailed account including the history of the development of the different variations of the Nice model see the recent paper by \citet{clement:18}. The hypothesis of the Nice model is that the giant planets formed in a much more compact configuration than their present day positions suggest. In this model the giant planets moved due to interactions with a remnant disc: due to conservation of momentum, the outer planets moved outwards, because of scattering objects from an exterior disk of bodies preferentially inwards. By contrast,  Jupiter moved in, because it was likely to eject small bodies from the system in this configuration. As one consequence of the scattering of disc objects, the hot Kuiper belt formed  with its objects moving on eccentric and inclined orbits  \citep{fernandez:84,hahn:99,thommes:99,gomes:03,tsinganis:05,morbi:05,levison:08,nesvorny:15}. The Nice model has changed significantly since
it was introduced, so that basically a entire family of models exists now, which differ in their initial conditions. They commonly reproduce the hot Kuiper belt population very well plus some additional features which vary from model to model.  

Originally the Nice model neither gave an explanation for the cold Kuiper belt population, nor predict the Sednoids, nor explain the higher mass of Neptune compared to Uranus,  nor necessarily end up with a ninth planet, nor the newly discovered populations mentioned above. However, by now there are potential explanations for each of these features, although not necessarily in the same model realization. In the early models the stability of the terrestrial planets was problematic, therefore
it was proposed that Jupiter jumped over its 2:1 mean motion resonance with Saturn. It turned out that by starting with a configuration that contains initially more giant planets the system becomes more stable.  This means by now, many realisations of the Nice model contain initially 5, 6 or even 7 giant planets. The fact that Neptune is more massive than Uranus is now explained by Neptune and Uranus initially having been positioned in opposite order than today and would have swapped places due to an instability  \citep{kaib:16}.  

The 10\% of the Kuiper belt objects that move on fairly circular, low-inclination orbits (referred to as cold Kuiper belt) is basically not achievable by a scattering process. Therefore it is now assumed that a cold Kuiper belt population was produced from a remnant population that was not scattered and located outside Neptune. When Neptune moved outwards this population was moved to more distant orbits, too. However, this only works if Neptune migrates outwards quite slowly \citep{parker:11,nesvorny:15}. Recently, an alternative scenario was suggested with a five planet configuration where the cold population formed in situ, with an outer edge between 44 - 45 AU, which never had a large mass \citep{gomes:18}.

The planets themselves could not have excited the orbits of the Sednoids, but an external forces is required. One recently revived suggestion for such an external force is the presence of an additional planet on a distant eccentric orbit. There are different possible origins of such a potential ninth planet or planet X. In context of the Nice model it has been suggested that one of the additional planets in the initial configuration was excited to this orbit. The newly discovered populations of TNOs mentioned above could be excited to their orbits within the Nice model through combined interactions with Neptune’s mean motion resonances and the Kozai resonance \citep{sheppard:16b}. Thus we see that there do exist explanation for all these phenomena within the Nice model. However, quite a few processes have to have happened in a certain order. 

Comparing the Nice model and the above described model is not straightforward for two reasons. First, the Nice model has first been suggested in 1984 and has been developed over 30 years, by now a large body of work exists improving the model was a gradual process over time.  Second, the Nice model consists not of a single set up that explains all these observed outer solar system features, but a family of set ups that have in common that they start with a compact configuration followed by planet migration afterwards.
So one has to keep in mind that this is a comparison between one single set up and an entire family of model realisations.

The current situation is that the Nice model seems to produce the hot Kuiper belt population in a long-term development. Note that some
authors suggest that so far most N-body simulations do not account for the effects of gravitational interactions between the disc particles
 itself  \citep{nesvorny:12}. This needs to be tested in the future to be sure that the Nice model reproduces the hot Kuiper population well. For the here presented model this long-term development has still to be demonstrated. So in terms of the hot Kuiper belt population the Nice model is currently better proven. However,  to find explanation for other outer solar system features it makes the model more complex and sets increasingly stringent criteria on the initial conditions. Thus in terms of simplicity the here presented fly-by scenario is at an advantage in comparison to the Nice model.


So far the strongest argument against the fly-by scenario has been the probability of such an event happening. As shown above, this no longer seems so problematic. To put  this into perspective we would have to compare it to the probability of the Nice model. This is very difficult to access as the exoplanet statistics is still too restricted in the relevant parameter range to give reliable estimates for the probability of finding an initial situation as demanded by the various realizations of the  Nice model. However, there have been some estimates for the development of such initial configurations to the current situation in the solar system. There seems to be a 
1\% chance that the terrestrial planets' orbits and the giant planets' orbits are reproduced simultaneously \citep{kaib:16}. The likelihood for Neptune and Uranus swapping places is, at least for the investigated parameter space, very high at 50\%\citep{levison:08}.  Thus the probability of such a realization would be 5 $\times$ 10$^{-3}$. Unfortunately there are no estimates on how often a sufficiently slow movement of Neptune is achieved necessary for retaining a cold Kuiper belt population.  However, there exist an estimate how likely it is that a ninth planet is excited on a suitable orbit, which then could lead to the orbits of the Sednoids. In about 5\% of simulations one of the additional planets is excited to a distant eccentric orbit as required for the ninth planet. Thus if we would assume that these probabilities are independent, we would obtain in 2.5 $\times$ 10$^{-4}$ of all cases a solar system equivalent. This estimate does neither include the likelihood for a slow movement of Neptune nor that for the initial configuration and therefore can only regarded as an upper limit. A more detailed calculations are definitely required for the future. However, taking it at face-value the argument against the fly-by scenario seems not a very severe one, because one could come up with a similar argument against, at least this realization, of the Nice model.

\subsection{Capture model}

As mentioned in the introduction an alternative suggestions for the origin of the Sednoids is the capture from the outer disc of another star during a close fly-by  \citep{morbi:04, kenyon:04,jilkova:15}. Here we compare our results to the most recent ones by  \citet{jilkova:15}. They performed a parameter study looking for a best fit between the particles
captured during such a fly-by and the Sednoids. They assumed that the Kuiper belt is formed according to the Nice mechanism and that the Sednoids were  added by capture during a fly-by. Therefore the investigated parameter space is restricted to fly-bys that leave the inner 50 AU undisturbed, which means they only investigate fly-bys with periastron distances $b>$ 265 AU and masses $M_2 <$ 2\Msun. They find the best match for the Sednoids for fly-bys of a 1.8 \Msun\ star that impacted the Sun at approximately 340 AU at an inclination 
with respect to the ecliptic of 17 to 34$^\circ$  with a relative velocity at infinity of $\approx$ 4.3 km/s. 

First we only look at the Sednoids and ask the question whether it is more likely that the Sednoids got captured than having originated from the solar system itself in such a fly-by scenario.
Actually the likelihood is about the same, the reason is the following: The Sednoids can be captured in a more distant fly-by at 340 AU compared to 100 AU, which makes such fly-bys more likely. However,  this is more than outbalanced by the fact that the star would have a more than 3 times higher mass. As a consequence of the IMF, fly-bys of more massive stars are less likely by about a factor of 20. This means that in total the chances for the Sednoids originating form the Sun's own disc are slightly higher than for being captured. This is independent from whether the fly-by happened within the first 10 Myr or considerable afterwards. When one combines this with the Nice model for the Kuiper belt, the combined probability is smaller than that for above presented model.

\subsection{Ninth planet}
 
Here we want to test whether the existence of a ninth planet would be consistent with a fly-by scenario. It is obvious from above that a potentially existing ninth planet orbiting on a roughly circular orbit, could be easily excited to more distant eccentric orbit due to a fly-by.  However, the question is where would that planet have resided before the fly-by. 
One suggestion for the rings in observed discs is that forming planets carve these rings into the discs. When one combines this idea, with a feature that neither the Nice nor the fly-by model explains, namely the low fraction of objects in the 50-70 AU range,  this would be a candidate location for the ninth planet before a fly-by.
 
From the extension of the surface density one would expect that at 50-70 AU a planet would have a mass of \mbox{1-2 $M_{Earth}$.} If a planet was located at 50-70 AU  and cleared this area we find a 10\% likelihood that it would have been ejected. Alternatively,  it  could have been flung onto a wide eccentricity orbit with a probability of 50\%: If it went on an extreme orbit with an eccentricity $>$ 0.8 this would happen in 5\% of all cases. Thus we would expect a lower mass for a potentially existing ninth planet. Alas such a low mass object would probably not have been strong enough to excite the orbits of the Sednoids, but it would be sufficient to stabilize them on their tracks. Thus a ninth planet and a fly-by scenario would be well compatible.  
    
\subsection{Hybrid models} 

As mentioned above the different scenarios discussed here are not necessarily mutually exclusive, but one could also imagine hybrid scenarios.
For the capture model, this would naturally result if there would be a close fly-by \mbox{($\approx$ 100 AU)} as described above, but the perturber would also be surrounded by a disc. 
In this case, there would be two population of TNOs, one originating for the solar system and one external. This would be a good explanation for the observed bi-modality in the properties of the TNOs in terms of color. In the future this scenario should be explored in more detail.

In the here presented model we did not consider the situation inside 35 AU in the solar system, because this area would be basically unaffected by the fly-bys considered here. Thus in principle, it is not necessary that the planets inside 35 AU were always at their current position, but they could have formed equally well in a compacter configuration.
Thus, a close fly-by and the Nice mechanism are not mutually exclusive. In this situation the Nice model could still account inner solar system features and contribute to the Kuiper belt population whereas the fly-by would be responsible for the Sednoids. In this situation the Kuiper belt would consist of a mixed population which again would be an explanation for the bi-modality of these objects.

\section{Summary and Conclusion}

Here we investigated the possibility that the orbits of the TNOs were caused by the fly-by of another star. Starting point of our simulations was the Sun surrounded by a disc, which could either be a protoplanetary or a debris disc, possibly containing already formed planets. We performed an extensive parameter study for the effect of close fly-bys on discs concentrating on the ones that lead to a steep drop in the mass distribution at 30-35 AU as that observed for the solar system. We found that fly-bys of stars with masses in the range 03.-1.0 \Msun\ at aperihelion distances of between 50 and 150 AU inclined  between 50 to 70 degree and at an angle of periastron between 60 to 120 degree are the most promising candidates for a fly-by shaping the outer solar system. Such fly-bys lead to Sednoids, a hot and cold Kuiper belt population and various other  properties characteristic for the outer solar system.  What distinguishes this model from others, is that only a single event is necessary to create all this features. Thus the beauty of this model lies in its simplicity. 

In a different set of simulations we determined the likelihood of such fly-bys over time. Here we choose a cluster similar to the ONC as model environment. 
Performing simulations of the dynamics of star clusters - gas expulsion, expansion and finding a new quasi-equilibrium - we find that such type of fly-bys are probable not only during the early phases but also on Gyr time scales, with a 5-7\% chance of all solar-type stars experiencing such a fly-by during the first 10 Myr and a 20-30\% chance in the next Gyr. If planet formation is as fast as anticipated from the ring structures in some discs, even a fly-by during the first 10 Myr would be an option. We showed that the probability of such an event even in the consecutive 4.56 Gyr is competitive to that of other models.

In summary, a close fly-by offers a realistic alternative to the present models suggested to explain the unexpected features of the outer solar system. It should be considered alongside these models as an option for shaping the outer solar system. The strength of this hypothesis lies in its simplicity by explaining several of the outer solar system features by one single mechanism. 
Thus this study should be regarded as a first proof-of-principle investigation showing that fly-bys in the parameter range defined above can produce many of the features seen in the solar system. The next steps would be to test within this parameter space for the ideal fit to all theouter solar system features and calculate how the Kuiper belt population would develop after a fly-by due to the interactions with Neptune and self-gravity within the disc.

Predictions from this model would be the following: 
\begin{itemize}
\item Objects like 2012 VP$_{ 113}$ should be more common than those with parameters like Sedna and preferentially located  at high inclinations.  
\item We find that in most close fly-bys not only features similar to the hot Kuiper belt, but also often Sednoid-like populations are produced though their actual location and population strength varies: Similar structures should also exist around many other planetary systems.
\item There should exist considerably more members of the new family TNOs found by  \cite{sheppard:16} at even large periastra.
\end{itemize}

\bibliographystyle{unsrt}

\end{document}